\documentclass{ws-ijmpa}
\usepackage{relsize}
\usepackage{amsmath}
\RequirePackage{xspace}
\def\babar{\mbox{\slshape B\kern-0.1em{\smaller A}\kern-0.1em
    B\kern-0.1em{\smaller A\kern-0.2em R}}}
\def\Bbar    {\kern 0.18em\overline{\kern -0.18em B}{}\xspace}

\def\BB      {\ensuremath{B\Bbar}\xspace}
\def\Bz      {\ensuremath{B^0}\xspace}
\def\Bzb     {\ensuremath{\Bbar^0}\xspace}
\def\BzBzb   {\ensuremath{\Bz {\kern -0.16em \Bzb}}\xspace}
\def\dE{{\Delta E}}
\def\mes{\ensuremath{m_{\rm ES}}}

\newcommand{\tev}{\ensuremath{\mathrm{\,Te\kern -0.1em V}}\xspace}
\newcommand{\gev}{\ensuremath{\mathrm{\,Ge\kern -0.1em V}}\xspace}
\newcommand{\mev}{\ensuremath{\mathrm{\,Me\kern -0.1em V}}\xspace}
\newcommand{\kev}{\ensuremath{\mathrm{\,ke\kern -0.1em V}}\xspace}
\newcommand{\ev}{\ensuremath{\mathrm{\,e\kern -0.1em V}}\xspace}
\newcommand{\gevc}{\ensuremath{{\mathrm{\,Ge\kern -0.1em V\!/}c}}\xspace}
\newcommand{\mevc}{\ensuremath{{\mathrm{\,Me\kern -0.1em V\!/}c}}\xspace}
\newcommand{\gevcc}{\ensuremath{{\mathrm{\,Ge\kern -0.1em V\!/}c^2}}\xspace}
\newcommand{\mevcc}{\ensuremath{{\mathrm{\,Me\kern -0.1em V\!/}c^2}}\xspace}
\begin{document}
\markboth{Haibo Li}
{Constraints on the CKM angle $\alpha$ in the $B \rightarrow \rho\rho$ decays}

%
\catchline{}{}{}{}{}
%
\title{Constraints on the CKM angle $\alpha$ in the $B \rightarrow \rho\rho$ decays}
\author{\footnotesize Haibo Li }
\address{Physics Department, University of Wisconsin, Madison,
Wisconsin 53706, USA }
\maketitle
\pub{Received (1 November 2004)}{}

\begin{abstract}
Using a data sample of 122 million $\Upsilon(4S) \rightarrow \BB$
decays collected with \babar~ detector at the PEP-II asymmetric $B$ factory
at SLAC, we measure the time-dependent-asymmetry
parameters of the longitudinally polarized component in the $B^0 \rightarrow
\rho^+\rho^-$ decay as $C_L = -0.23 \pm
0.24 ({\rm stat}) \pm 0.14 ({\rm syst})$ and $S_L = -0.19 \pm
0.33 ({\rm stat}) \pm 0.11 ({\rm syst})$. The $B^0 \rightarrow \rho^0\rho^0$ decay mode
is also searched for in a data sample of about 227 million $B\overline{B}$ pairs.
No significant signal is observed, and an upper limit of $1.1\times 10^{-6}$
(90\% C.L.) on the branching fraction is set. The penguin contribution to
the CKM angle $\alpha$ uncertainty is measured to be 11$^o$. All results are
preliminary.

\end{abstract}

\section{Introduction}	

The time-dependent $CP$ asymmetry in a $b
\rightarrow u\bar{u}d$ decay of a $B^0$ to a $CP$ eigenstate allows for
a measurement of the angle $\alpha = arg[-V_{\rm td}V^*_{\rm tb}/V_{\rm
ud}V^*_{\rm ub}]$ if
the decay is dominated by the tree amplitude.  The
contribution from penguin diagrams gives rise to a correction $\Delta \alpha
= \alpha_{\rm eff} - \alpha$ that can be inferred through an isospin
analysis~\cite{iso}. The recent experimental results~\cite{charge} indicate a small
penguin contributions in $B \rightarrow \rho\rho$.
The $CP$ analysis in $B \rightarrow \rho^+\rho^-$ is complicated by the
presence of three helicity states.  However, the measured polarizations in
$\rho^+\rho^-$ and $\rho^+\rho^0$ modes indicate a dominance of the helicity
0 state (longitudinal polarization), that is a $CP=+1$ eigenstate.
A measurement of the polarization in $B^0 \rightarrow \rho^0\rho^0$ would
complete the isospin triangle, but this mode has not been observed so for.
Knowledge of the $B^0 \rightarrow \rho^0\rho^0$ rate
is still expected to be limiting factor to the accuracy of the $\alpha$ measurement with $\rho\rho$
decays.

In this paper, we present a time-dependent analysis of $B^0 \rightarrow
\rho^+\rho^-$ based on a sample of 122 million \BB pairs, and a search for
the $\rho^0\rho^0$ final state on a sample of 227 million \BB pairs at
\babar.
%
\section{Analysis Method}

We reconstruct $\rho^+\rho^-$
candidates from combinations of two charged tracks and two
$\pi^0$ candidates. In the $\rho^0\rho^0$ mode,
the $B^0$ candidates are reconstructed from their decay products $\rho^0
\rightarrow \pi^+\pi^-$ with four charged tracks which are required to originate from a
single vertex near the interaction point.
The $\pi^0$ candidates are formed from pairs of photons that have measured
energies greater than 50 MeV. The reconstructed $\pi^0$ mass must satisfy
$0.10 < m_{\gamma\gamma} < 0.16 $ GeV/c$^2$. The mass of the $\rho$
candidates, $m_{\pi^{\pm}\pi^0}$, must satisfy $|m_{\pi^{\pm}\pi^0} - 0.770
\gevcc| < 0.375$ $\gevcc$, and the mass, $m_{\pi^+\pi^-}$, must
satisfy $0.55<m_{\pi^+\pi^-}<1.0$ $\gevcc$.
Combinatorial backgrounds dominate near $|{\rm cos}\theta_i| = 1$
, where $\theta_i$, $i=1$, 2 is defined for each $\rho$ meson as the
angle between the $\pi^0$ ($\pi^+$) momentum in the $\rho^{\pm}$ ($\rho^0$) rest
frame and the flight direction of the $B^0$ in this frame,
We reduce these backgrounds with the requirement $-0.8 <
{\rm cos}\theta_i < 0.98$ in $\rho^+\rho^-$ modes and $|{\rm cos}\theta_i| < 0.99$
in $\rho^0\rho^0$ mode.
Two kinematic variables~\cite{mes}, $\dE$ and $\mes$, allow the discrimination
of signal $B$ decays from random combinations of tracks and
$\pi^0$ candidates.
For $\rho^+\rho^-$ we require that $5.21<\mes<5.29$
$\gevcc$ and $-0.12<\dE<0.15$ $\gevcc$. The asymmetric $\dE$ window
suppresses background from higher-multiplicity $B$ decays. For
$\rho^0\rho^0$ we require $\mes>5.24$ $\gevcc$ and $|\dE|<85$ $\mevcc$.

In order to reject the dominated quark-antiquark continuum background, we
require $|{\rm cos}\theta_T|<0.8$, where $\theta_T$ is the the angle between the
$B$ thrust axis and the thrust axis of the rest of the events (ROE).
The other event-shape discriminating variables are combined in a neural network~\cite{nn}
(${\cal N}$).
The ${\cal N}$s for $\rho^+\rho^-$ and
$\rho^0\rho^0$ analysis weight the discriminating variables differently,
according to training on off-resonance data and the relevant Mote Carlo
(MC) simulated signal events.

When multiple $B$ candidates can be formed we
select the one that minimizes the sum of the deviations of the reconstructed
$\pi^0$ mass in $\rho^+\rho^-$ mode, while, for $\rho^0\rho^0$, one
candidate is selected randomly. The selection efficiency is 7\% (13\%) for
the longitudinaly (transversely) polarized $\rho^+\rho^+$ signal,
and it is 27\% (32\%) for the $\rho^0\rho^0$ signal.
$B$ in the event.

To study the time-dependent asymmetry one needs to measure the proper time
difference, $\Delta t$, between the two $B$ decays in the events, and to
determine the flavor tag of the other $B$-meson. The time difference between
the decays of the two neutral $B$ mesons ($B_{\rm rec}$, $B_{\rm tag}$) is calculated
from the measured separation $\Delta z$ between the $B_{rec}$ and $B_{tag}$
decay vertices~\cite{babarbrec}. The flavor of the $B_{\rm tag}$ is determined
with a multivariate technique~\cite{mes} that has a total effective tagging
efficiency of (28.4$\pm$0.7)\%.

An unbinned likelihood fit is finally performed on the selected event, a
probability density function is built from discriminating variables,
including the angular distribution and the $\Delta t$-dependence.
The signal decay-rate distribution $f_+$ ($f_-$) for $B_{\rm tag} = B^0$
($\Bzb$) is given by:
\begin{equation}
f_{\pm} (\Delta t) =\frac{e^{-|\Delta t|/\tau}}{4\tau}[1\pm S_L
{\rm sin}(\Delta m_d \Delta t) \mp C_L {\rm cos}(\Delta m_d \Delta t)],
\label{time}
\end{equation}
where $\tau$ is the mean $B^0$ lifetime, $\Delta m_d$ is the
$B^0-\Bzb$ mixing frequency, and $S_L$ and $C_L$ are the $CP$
asymmetry parameters for the longitudinal polarized signal.

\section{Results}

We measure the $CP$ violating asymmetries in the $B^0 \rightarrow \rho^+\rho^-$
longitudinal component decay on 122 million \BzBzb pairs. A detailed analysis of
the background due to other $B$ decays is performed. The main systematic
uncertainty on the asymmetries is found to be the unknown $CP$ violation in
$B$ background events. Our results are $S_L = -0.19 \pm 0.33 ({\rm stat}) \pm
0.11({\rm syst})$ and $C_L = -0.23 \pm 0.24 ({\rm stat}) \pm 0.14 ({\rm syst})$.
With a sample of 227 million \BzBzb pairs we have searched for the decay
mode $B^0 \rightarrow \rho^0\rho^0$, the measured value for the branching
fraction is $(0.54^{+0.36}_{-0.32} \pm 0.19) \times 10^{-6}$ or
an upper limit of $1.1\times 10^{-6}$ at 90\% confidence level (C.L.).

\begin{figure}
\centerline{\psfig{file=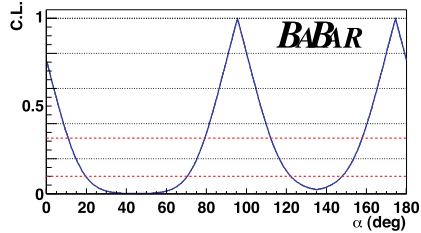,width=6cm}}
\vspace*{2pt}
\caption{Confidence Level on $\alpha$ (solid curve) obtained for this
results. The red dashed lines correspond to the 68\% (top) and 90\% (bottom)
confidence intervals.}
\label{alpha}
\end{figure}

Using the Grossman-Quinn bound~\cite{iso,quinn} with the recent results~\cite{charge} on
$B^{\pm} \rightarrow \rho^{\pm}\rho^0$ we limit
$|\alpha_{\rm eff}-\alpha|<11^o$ (68\% C.L.).
Ignoring possible non-resonant
contributions, and $I = 1$ amplitudes~\cite{one} one can
relate $CP$ parameters $S_L$ and $C_L$ to $\alpha$, up to a four-fold
ambiguity. If we select the solution closest to the CKM best fit central
value~\cite{ckm}, with the new limit on the $B^0\rightarrow \rho^0\rho^0$
rate we improve the constraint on $\alpha$ due to the penguin contribution,
the measured $CP$ parameters of the longitudinal
polarization correspond to $\alpha = (96 \pm 10({\rm stat}) \pm 4({\rm syst}) \pm 11
({\rm penguin}))^o$. Figure~\ref{alpha} shows the confidence level as a function
of $\alpha$ from the isospin analysis.
%

\end{document}